\documentclass[%
aps,
 reprint,
 amsmath,amssymb,
prper,
floatfix
]{revtex4-2}

\usepackage{graphicx}
\usepackage{dcolumn}
\usepackage{bm}
\usepackage{tikz}
\usepackage{pgfplots}
\usepackage[export]{adjustbox}
\pgfplotsset{compat=1.17}

\newcommand{\attr}[1]{\textit{#1}}

\begin{document}


\title{Cheat sites and artificial intelligence usage in online introductory physics courses:\\ 
what is the extent and what effect does it have on assessments?}

\author{Gerd Kortemeyer}
 \email{kgerd@ethz.ch}
 \affiliation{%
Rectorate and AI Center, ETH Zurich, 8092 Zurich, Switzerland
}%
\altaffiliation[also at ]{Michigan State University, East Lansing, MI 48823, USA}

\author{Wolfgang Bauer}
 \email{bauerw@msu.edu}
 \affiliation{%
Department of Physics and Astronomy, Michigan State University, East Lansing, MI 48824, USA
}%

\date{\today}

\begin{abstract}
As a result of the pandemic, many physics courses moved online. Alongside, the popularity of internet-based problem-solving sites and forums rose. With the emergence of Large Language Models, another shift occurred. One year into the public availability of these models, how has online help-seeking behavior among introductory physics students changed, and what is the effect of different patterns of online-resource usage? In a mixed-method approach, we investigate student choices and their impact on assessment components of an online introductory physics course for scientists and engineers.  We find that students still mostly rely on traditional internet resources, and that their usage strongly influences the outcome of low-stake unsupervised quizzes. However, we also find that the impact of different help-seeking patterns on the supervised assessment components of the course is non-significant.
\end{abstract}

\maketitle

\section{Introduction}
The general assumption in teaching introductory physics courses is that we need to teach a few essential concepts, for example conservation laws, and, based on those, to derive basic equations that govern the motion of all objects in our universe, from atomic nuclei to galaxies.  Almost all of us teaching professionals subscribe to the notion that it aids the students' learning processes to flesh out these basic concepts with exercises, in class and as homework.  Mastering our subject requires time spent on wrestling with conceptual questions, and familiarity with basic equations is acquired by rearranging and combining them and solving for the unknown quantities.

``I think, however, that there isn't any solution to this problem of education other than to realize that the best teaching can be done only when there is a direct individual relationship between a student and a good teacher--a situation in which the student discusses the ideas, thinks about the things, and talks about the things. It's impossible to learn very much by simply sitting in a lecture, or even by simply doing problems that are assigned. But in our modern times we have so many students to teach that we have to try to find some substitute for the ideal.''  These sentences were written by Richard Feynman.  In 1963~\cite{feynman1965m}!

Some of us, the authors included~\cite{kortemeyer1999,bauer2023}, have spent enormous amounts of time to construct learning management systems, audience feedback systems, simulation software, and other apps to help with giving the students an opportunity to spend meaningful time on task.  The ultimate goal may have been to construct a software system that acts as a personal tutor for each individual student.  The rapid rise of Large Language Models (LLMs) and artificial intelligence (AI) systems gives hope that the substitute for the ideal that Feynman described might become reality.  Alternatively, students might just use their new AI tools to sidestep their learning process and just let the AI complete the assignments for them.

For the last 25~year, Michigan State University has been running online sections of the introductory physics courses~\cite{kortemeyer2014onl}. For decades, even preceding online courses, the prevalent notion regarding course delivery modes had been ``no significant difference''~\cite{russell1997no,cavanaugh2015large,bergeler2021comparing}. More recently, we confirmed this for the courses under investigation in this study when we found no significant difference between attendance choices with regards to learning success~\cite{kortemeyer22hybrid} and preparation for subsequent courses~\cite{kortemeyer2023taking}.

\begin{figure}
\begin{center}
\includegraphics[width=\columnwidth]{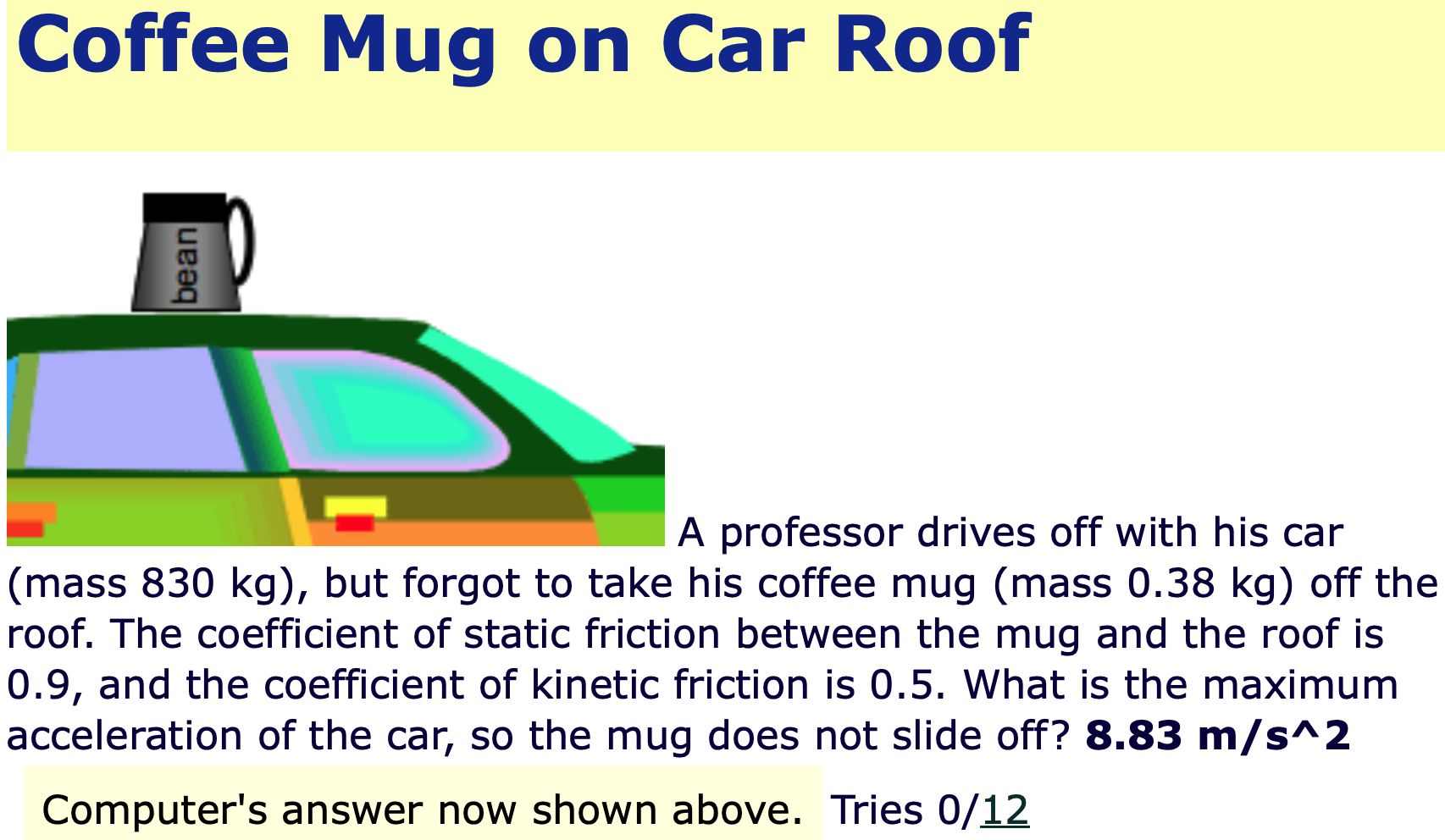}

\includegraphics[width=\columnwidth]{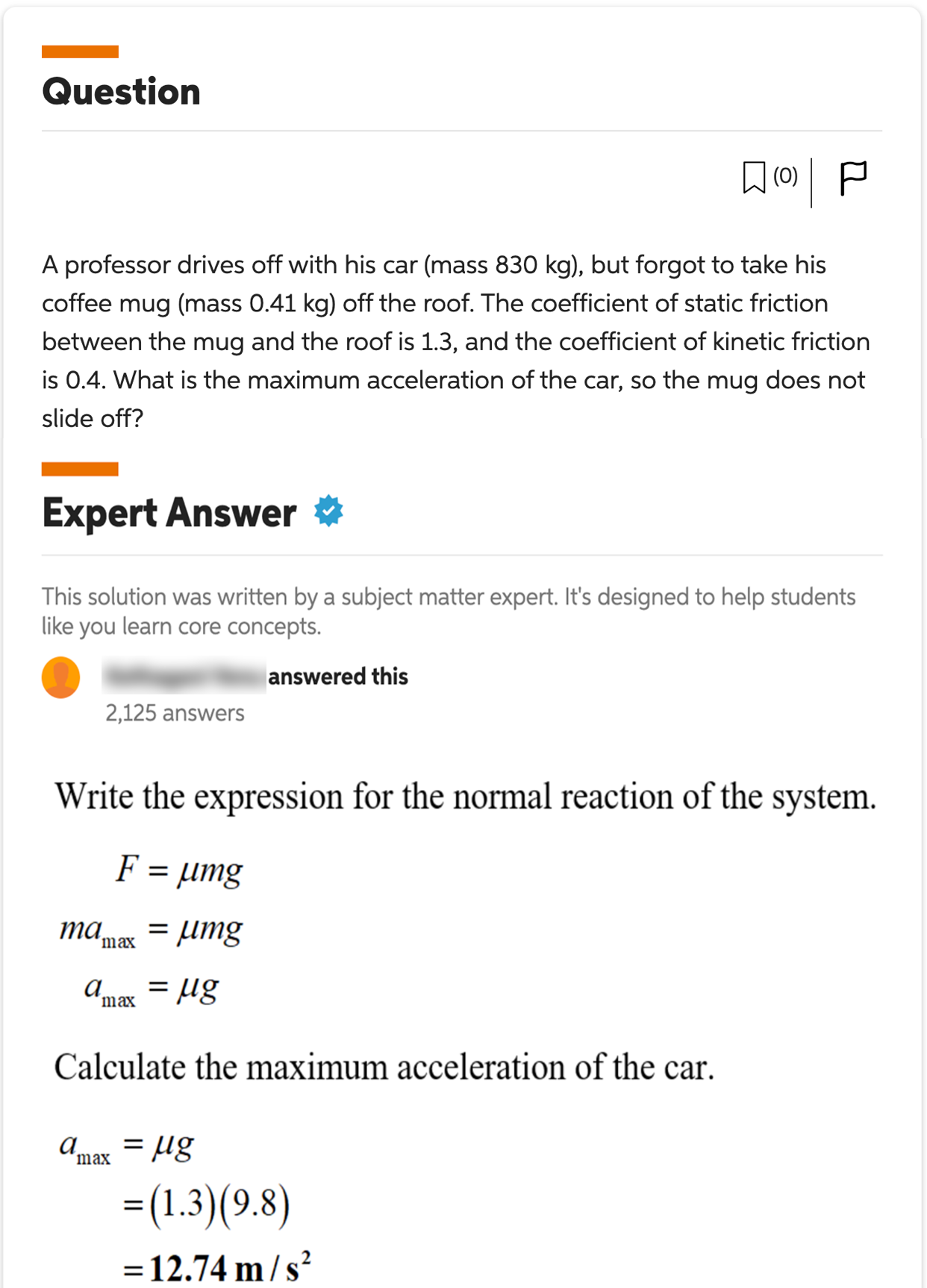}
\end{center}
\caption{A problem written and copyrighted by one of the authors (GK) in LON-CAPA~\cite{kortemeyer08} (top panel) and a typical solution found on a problem-solving site (bottom panel).  The problem numbers are randomized, so the student would need to identify their own numerical values and insert them. Depicted is one of 50~answers found on the site.\label{fig:forum}}
\end{figure}

Since the courses first came online, the amount of additional resources available online has greatly increased, particularly during COVID-19. The solution to virtually any introductory-physics problem is available shortly after it is published~\cite{ruggieri2020students}, including on commercial sites like Chegg~\cite{chegg,lancaster2021contract}; a typical example can be seen in Fig.~\ref{fig:forum}. Students find these resources useful for homework or unsupervised online exam problems that have been ``recycled'' from earlier in the semester (e.g., homework problems making an encore appearance on exams).

Instructors have been fighting these internet sites and forums by editing the problem content, removing any references to problem numbers, randomizing the solutions, selecting problems from older editions of textbooks, or writing new problems every time~\cite{busch2017one}. All of these counter measures rely on there being some time delay between publishing a problem and it appearing online (even contracted problem solvers need a little time) and on the sites having one static version of the problem (e.g., not being able to adapt to randomizations).

\begin{figure*}
\begin{center}
\includegraphics[valign=t,width=\columnwidth]{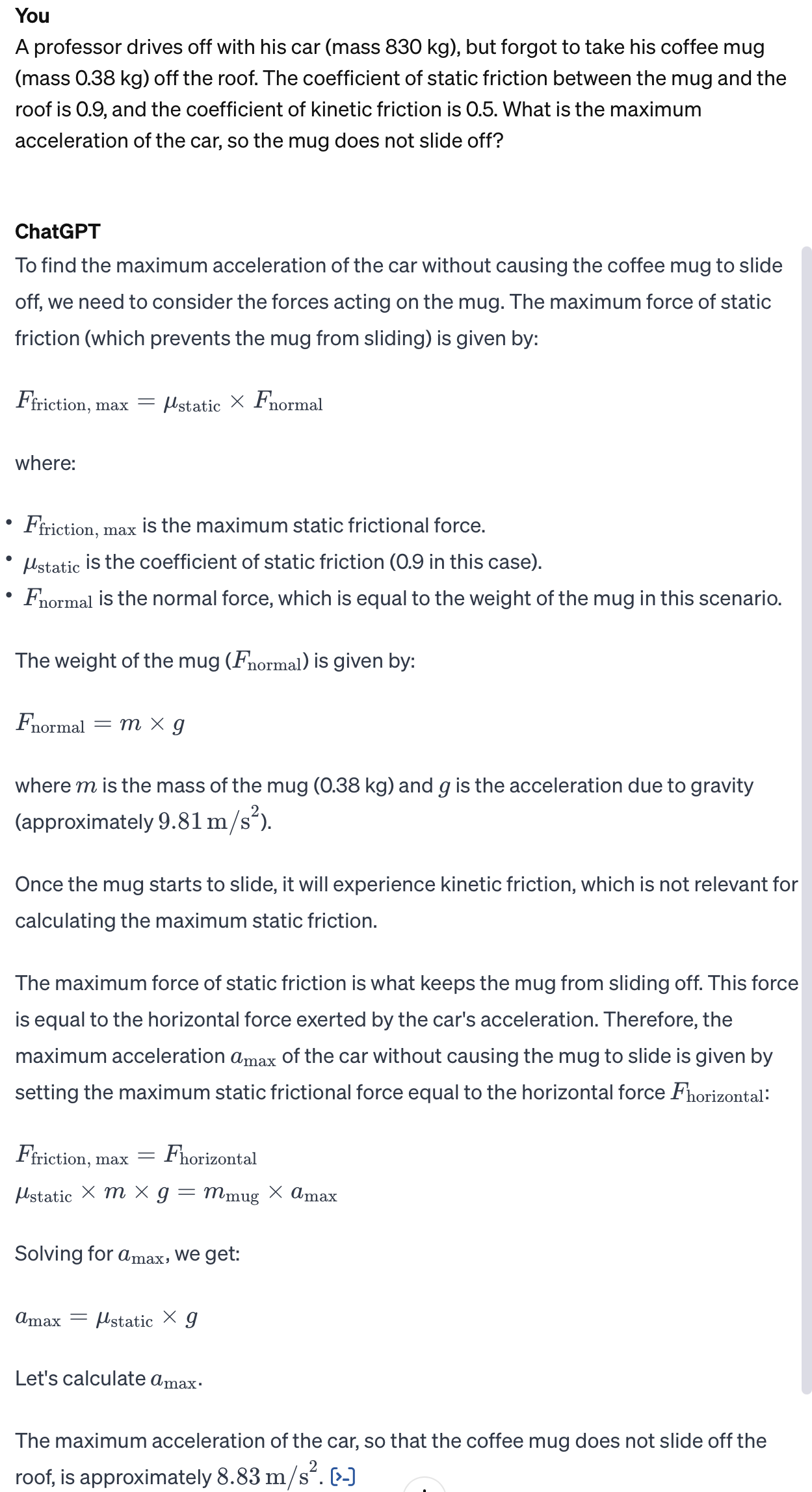}
\includegraphics[valign=t,width=\columnwidth]{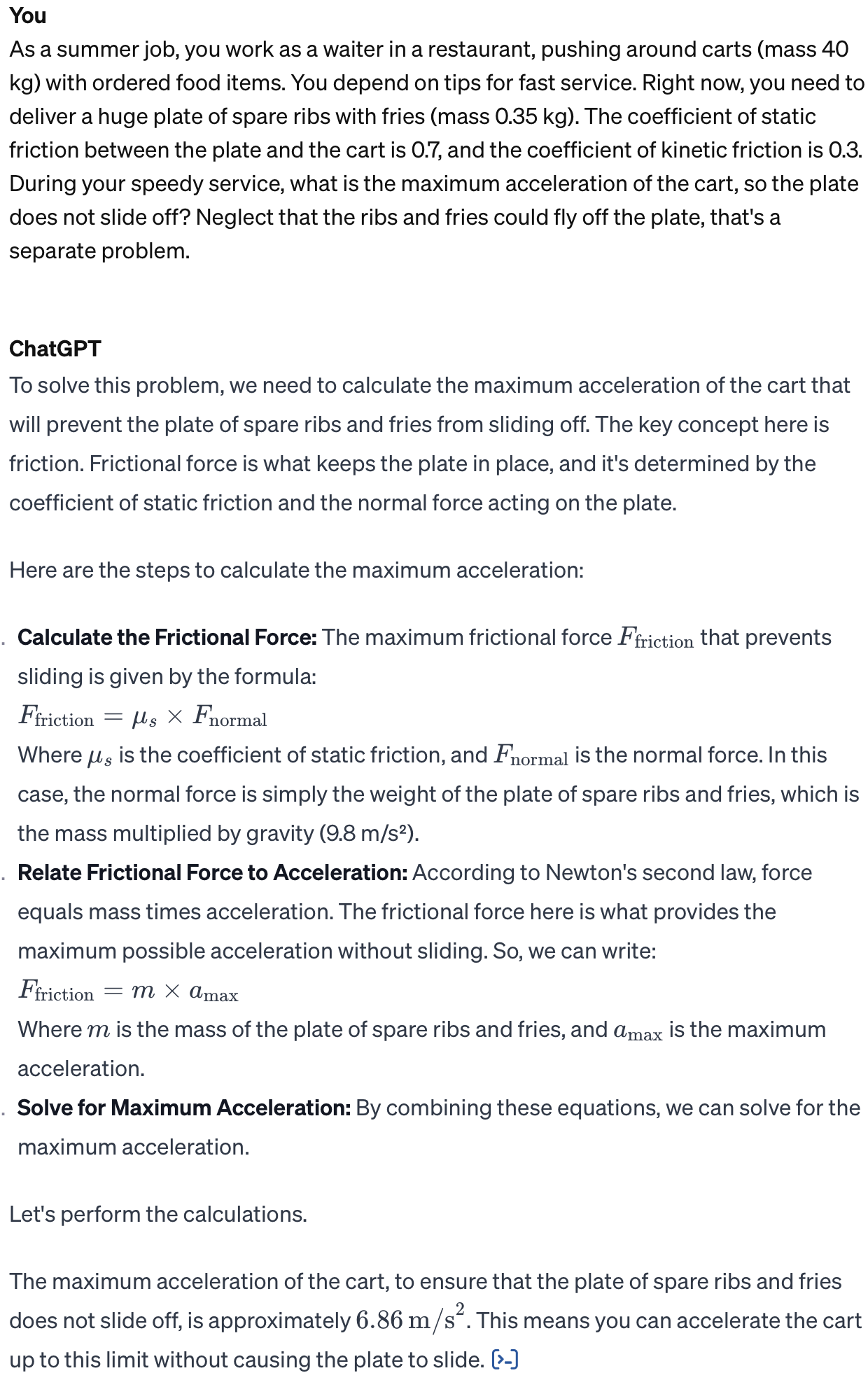}
\end{center}
\caption{The problem from Fig.~\ref{fig:forum} in its original form (left panel) and after on-the-fly modification (right panel), correctly solved by GPT-4 on the first attempt.}
\label{fig:gpt}
\end{figure*} 

These same counter measures against cheat sites will not work against artificial intelligence.  Tools like ChatGPT~\cite{chatgpt}  and Bard~\cite{bard} deliver solutions {\it ad-hoc} and on-demand, they are immune against changing wording or numbers, and they solve problems independent of them having been published days ago or appearing for the first time on an online exam. This is illustrated in Fig.~\ref{fig:gpt}, where the original friction problem has been modified by introducing additional distractors and different randomized numbers. Not only does the availability of these solutions depend on some ``expert'' having solved the problem before, but as opposed to the forum answer in Fig.~\ref{fig:forum}, the response includes the actual numbers encountered by the student and the physics explanation is arguably better and more helpful.

\begin{figure}
\begin{center}
\includegraphics[width=\columnwidth]{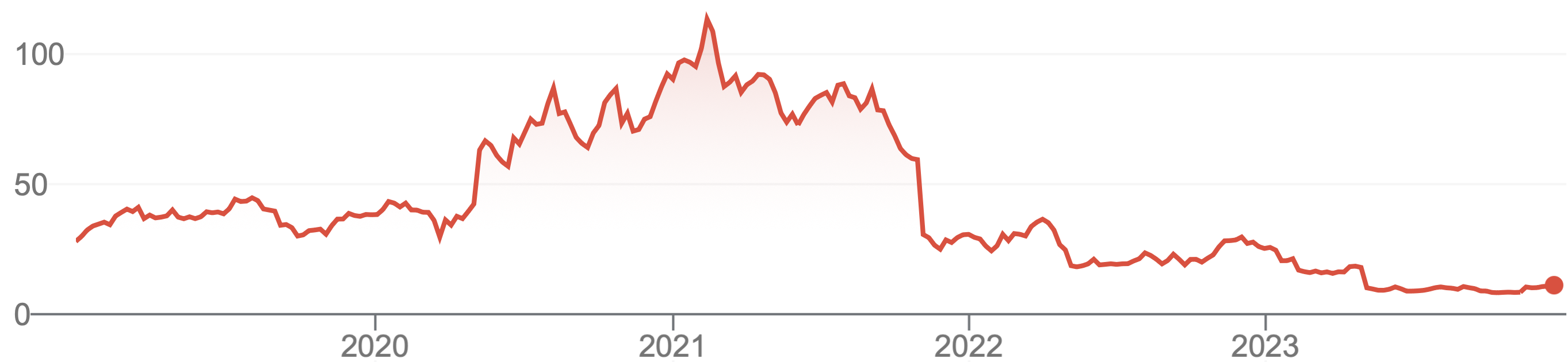}
\end{center}
\caption{Historic share prices in USD of Chegg (NYSE CHGG) as a proxy for the popularity of online problem-solving sites. Prices greatly increased when courses went online as a result of the pandemic, but fell again dramatically coinciding with the emergence of Large Language Models. \label{fig:chgg}}
\end{figure}

Chegg can serve as a proxy to the popularity of online problem-solving sites: while the share price of Chegg (NYSE: CHGG) greatly increased when courses went online due to the onset of the pandemic in 2020, with the appearance of ChatGPT in late 2022, they dropped below pre-pandemic levels~\cite{forbes,wired}, see Fig.~\ref{fig:chgg}.

Physics may still have been spared from this, since Large Language Models as ``calculators for words'' are still notoriously bad at math, and other online resources may still be more reliable. However, it has been shown that even older versions of popular AI~chatbots can (barely) pass the assessment components of introductory physics courses~\cite{kortemeyer23ai}, and as Fig.~\ref{fig:gpt} shows, newer versions make less calculation errors.

In our study, we investigate if one year into the public availability of powerful Large Language Models, online help-seeking behavior of students in an introductory physics course has shifted from traditional resources to AI. We also investigate the possible impact of different help-seeking patterns on the assessment components of such a course.

\section{Setting}
Michigan State University is a public, large-enrollment ($>50,000$ students) R-1 university. Almost 78\% of the undergraduate population are from Michigan. The online courses in this study were taught asynchronously using a variety of multimedia components~\cite{kortemeyer2014onl}. 

The study takes place in a calculus-based introductory physics course sequence for scientists and engineers, where both a first-semester mechanics and a second-semester E\&M course were offered during Fall semester 2023. Each course offered several asynchronous video lessons every week and online homework using LON-CAPA~\cite{kortemeyer08}. The courses had 11~low-stakes weekly exams (``quizzes'')~\cite{laverty12b}, of which nine were conducted online and two of which had to be taken on-campus under supervision. Faculty sanctioned the use of the textbook and the LON-CAPA materials during these exams, no other resources were allowed..

The course also included a high-stakes  on-campus final exam. The final exams included five questions which were randomized duplicates of problems had been assigned earlier in the semester. As a resource for the students, the course also offered an on-campus and online help room staffed by course faculty and staff.

At the end of the semester, a survey was given asking students to report how frequently they consulted artificial intelligence tools and other online resources during homework and online quizzes, and how often they conversed with fellow students and course faculty and staff while working on homework.
\section{Methodology}
\subsection{Survey administration}
The survey contained the following items, which for the numerical answers had sliders ranging from 0--100\%:
\begin{itemize}
\item Homework:
Estimate the percentage of your homework and lecture problems, for which the following is true:
\begin{itemize}
\item You used AI tools like ChatGPT, Khanmigo, \ldots to solve them (\attr{HwkAI}).
\item You used Internet resources like help sites or forums to solve them (\attr{HwkInt}).
\item You consulted other student to solve them (\attr{HwkPeer}).
\item You consulted the TAs/prof to solve them (\attr{HwkFac}).
\end{itemize}
\item Online Exams: Estimate the percentage of your online exam problems, for which the following is true:
\begin{itemize}
\item You used AI tool to solve them (\attr{OnlAI}).
\item You used Internet resources like help sites or forums to solve them (\attr{OnlInt}).
\end{itemize}
\item Your Opinion:
Please tell us what you think about using AI in online classes; what should ideally be done; what should not be done?
\end{itemize}

The survey was administered online during the last week of the semester, but results were not viewed or analyzed until the grades for the course had been turned in. A nominal participation credit was given for submitting the survey, regardless of whether or not the students agreed to be part of the study. The students were aware of this protocol as part of the informed consent, and data was only analyzed for students who agreed to participate. The study was approved under MSU-IRB-STUDY00009987.
\subsection{Considered variables}
We compiled a range of variables that capture various aspects of student performance and behavior, shown in Table~\ref{tab:variables}. Key performance metrics include \attr{Hwk} (homework score), \attr{OnlExams} (score from online exams), \attr{CamExams} (score from on-campus exams with supervision), and \attr{Final} (score from the final exam, also conducted on-campus with supervision). These scores are presented as percentages, reflecting the students' achievement in each respective assessment. Additionally, \attr{Sem5} represents the scores for five problems initially available online, \attr{Final5} for the same problems when included in the final exam, and \attr{Diff5} indicating the score difference between these two settings. Thus, \attr{Diff5} can be used as a proxy for retention of concepts between the semester and the final exam. Finally, \attr{DiffExams} quantifies the score difference between online and on-campus quizzes, offering insight into performance variations across different assessment environments.

The dataset also encompasses variables related to the use of digital resources and student interactions. \attr{HwkAI} and \attr{OnlAI} denote the self-reported percentage of problems for which artificial intelligence (AI) tools were used during homework and online quizzes, respectively. Similarly, \attr{HwkInt} and \attr{OnlInt} represent the usage of other internet resources in these settings. Finally, \attr{HwkPeer} and \attr{HwkFac} quantify the self-reported extent of peer discussions and interactions with faculty during homework.
\begin{table*}
\caption{Summary of variables in the dataset\label{tab:variables}}
\begin{ruledtabular}
\begin{tabular}{llrrrr}
Variable & Description & Min & Max & Mean & Std. Deviation \\
\hline
\attr{CamExams} & On-Campus Exam Score (\%) & 0.00 & 100.00 & 53.80 & 22.73 \\
\attr{Diff5} & Difference in Online/Final Duplicate Problem Scores & -100.00 & 40.00 & -39.00 & 33.80 \\
\attr{DiffExams} & Difference in Online/On-Campus Exam Scores & -81.55 & 23.57 & -29.30 & 20.04 \\
\attr{Final} & Final Exam Score (\%) & 0.00 & 100.00 & 49.19 & 27.64 \\
\attr{Final5} & Final Exam Duplicate Problem Score (\%) & 0.00 & 100.00 & 51.22 & 33.37 \\
\attr{Hwk} & Homework Score (\%) & 25.00 & 100.00 & 86.60 & 15.68 \\
\attr{HwkAI} & AI Usage During Homework (\%) & 0.00 & 100.00 & 17.08 & 23.94 \\
\attr{HwkFac} & Discussions with Faculty During Homework (\%) & 0.00 & 100.00 & 16.32 & 24.29 \\
\attr{HwkInt} & Internet Usage During Homework (\%) & 0.00 & 100.00 & 50.00 & 30.18 \\
\attr{HwkPeer} & Peer Discussions During Homework (\%) & 0.00 & 100.00 & 32.79 & 31.44 \\
\attr{OnlAI} & AI Usage During Online Exams (\%) & 0.00 & 88.00 & 8.07 & 17.54 \\
\attr{OnlExams} & Online Exam Score (\%) & 43.69 & 98.06 & 83.10 & 11.77 \\
\attr{OnlInt} & Internet Usage During Online Exams (\%) & 0.00 & 100.00 & 23.38 & 29.39 \\
\attr{Sem5} & Online Problem Score (\%) & 20.00 & 100.00 & 90.23 & 17.33 \\
\end{tabular}
\end{ruledtabular}
\end{table*}

\subsection{Statistical methods}
Data were downloaded from the course management system, and survey results were merged using Python scripts. Calculations for this project were carried out using ChatGPT-4 Advanced Data Analysis~\cite{chatgpt} and R~\cite{rproject} (in particular qgraph~\cite{qgraph} and CTT~\cite{ctt}).

\section{Results}
\subsection{Response rate}
The first and second semester courses were completed by 156~and 183~students, respectively. Of these, 90~and 131~students agreed to participate in the study, bringing the total to 221~participants.

\subsection{Online versus on-campus exams}
\begin{figure}
\begin{tikzpicture}
\begin{axis}[
    ybar,
    bar width=7pt,
    xlabel={Score Range},
    ylabel={Number of Students},
    xmin=0, xmax=100,
    ymin=0, ymax=100,  
    xtick={5, 15, 25, 35, 45, 55, 65, 75, 85, 95},
    xticklabels={0-10, 10-20, 20-30, 30-40, 40-50, 50-60, 60-70, 70-80, 80-90, 90-100},
    xticklabel style={rotate=45, anchor=east},
    nodes near coords,
    nodes near coords align={vertical},
    legend pos=north west,
    ]
\addplot coordinates {
    (5, 5) (15, 7) (25, 20) (35, 24) (45, 27) (55, 36) (65, 37) (75, 29) (85, 23) (95, 13)
};
\addplot coordinates {
    (5, 0) (15, 0) (25, 0) (35, 0) (45, 3) (55, 10) (65, 20) (75, 32) (85, 75) (95, 81)
};
\legend{CamExams, OnlExams}
\end{axis}
\end{tikzpicture}
\caption{Comparison between score distributions for the nine low-stakes exams that were conducted online (red) and the two low-stakes exams that were conducted on-campus under supervision (blue).\label{fig:scoredist}}
\end{figure}
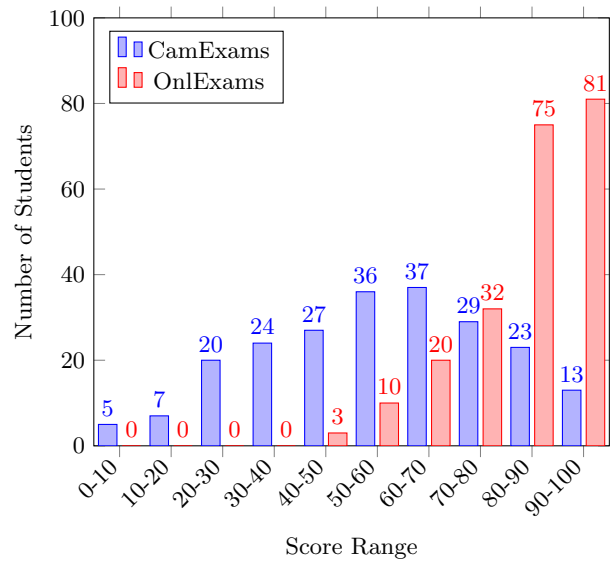

Figure~\ref{fig:scoredist} shows the score distributions for the nine exams that were conducted online and the two exams that were conducted on-campus under supervision. In a $t$-test, these distributions are significantly different ($p\approx4.7\cdot10^{-47}$).

The conditions under which these exams were conducted led to vastly different outcomes, and an immediate assumption would be that this is related to the use of external resources during unsupervised assessments. As an example, for nine of the ten questions on the last online exam, solutions could be found on Chegg within about one minute each.

\subsection{Usage of resources during unsupervised assessments}
\begin{figure}
\begin{tikzpicture}
\begin{axis}[
    ybar,
    bar width=20pt,
    xlabel={Variables},
    ylabel={Mean Value},
    symbolic x coords={OnlAI, OnlInt, HwkAI, HwkInt, HwkPeer, HwkFac},
    xtick=data,
    ymin=0,
    error bars/y dir=both,
    error bars/y explicit,
    nodes near coords,
    nodes near coords align={vertical},
    xticklabel style={rotate=45, anchor=east},
    ]
\addplot+[error bars/.cd, y dir=both, y explicit] coordinates {
    (OnlAI, 8.07) +- (0, 17.54)
    (OnlInt, 23.38) +- (0, 29.39)
    (HwkAI, 17.08) +- (0, 23.94)
    (HwkInt, 50.00) +- (0, 30.18)
    (HwkPeer, 32.79) +- (0, 31.44)
    (HwkFac, 16.32) +- (0, 24.29)
};
\end{axis}
\end{tikzpicture}
\caption{Self-reported usage of resources while working on homework and online exams (see Table~\ref{tab:variables} for the explanation of the variables).\label{fig:usage}}
\end{figure}
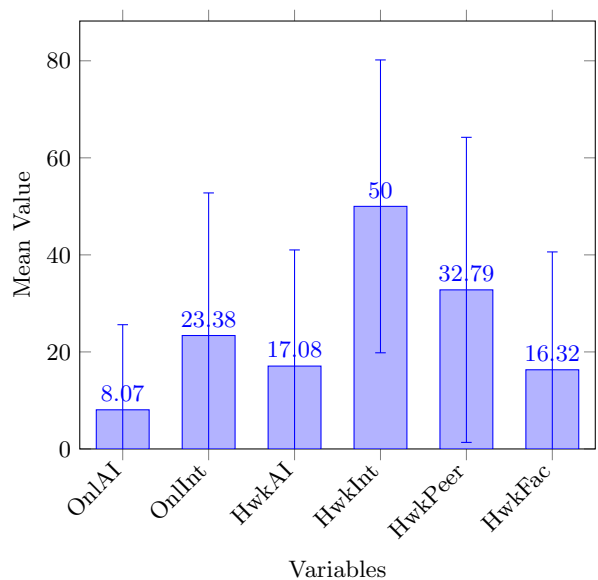

Figure~\ref{fig:usage} illustrates the average self-reported usage of AI and other internet resources, as well as self-reported consultation with peers and course personnel. 

\begin{figure*}
\begin{center}
\includegraphics[width=0.86\textwidth]{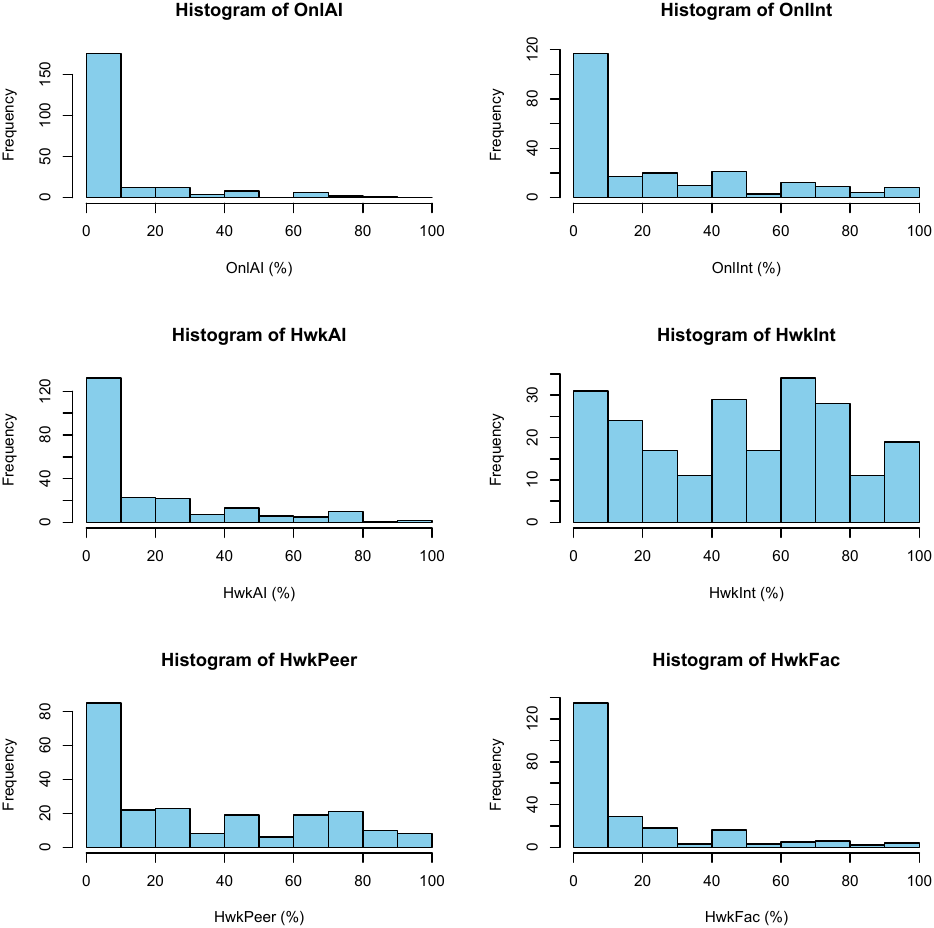}
\end{center}
\caption{Distribution of the survey responses.\label{fig:histo}}
\end{figure*}

Overall, students report less usage of resources during exams than during homework, but not significantly. The only one-sigma significant differences are between usage of AI and talking to faculty on the one hand, and using other internet resources while working on homework; students have not yet adopted AI and stick with ``traditional'' problem-solving sites. On the average, students use other internet resources for half of the homework problems. The distributions of these types of resources usages, however, are very different, see Fig.~\ref{fig:histo}, which suggests that there a different classes of resource usage.

Using $k$-means clustering and elbow method, we identified four different classes as indicated in Table~\ref{tab:clusters}. Cluster~1, the smallest group, is comprised of students who appear to prefer human interaction to any online resources, and these students mostly adhere to rules for the online exams. Students in Cluster~2, the largest cluster, state that they make little use of resources overall, and that they most closely adhere to rules for the online exams. Students in Cluster~3 make heavy use of  internet resources other than AI in both homework and exams, thus not following rules. Finally, students in Cluster~4 use all available resources and disregard exam rules.

\begin{table*}
\caption{Number of members and mean values of variables in identified clusters of resource usage.\label{tab:clusters}}
\begin{ruledtabular}
\begin{tabular}{llrrrrrrl}

Cluster&\# members&	\attr{HwkAI}&	\attr{HwkInt}&	\attr{HwkPeer}	&\attr{HwkFac}	&\attr{OnlAI}&	\attr{OnlInt}&Interpretation\\
\hline
1	&27&9.6&	39.9&	53.1&	66.4&	3.6	&14.3&Using mostly peer discussions and helprooms\\
2	&96&8.3	&29.3&	20.6&	5.7	&1.2&	3.1&Using resources on homework but not exams\\
3	&56&5.1	&80.1&	35.8&	7.4&	1.9&	51.3&Heavily using internet on homework and exams\\
4	&42&58.0&	63.7&	43.6	&20.3	&34.8&	38.2&Using all resources including AI everywhere
\end{tabular}
\end{ruledtabular}
\end{table*}

\subsection{Correlations of attributes}
\begin{figure}
\begin{center}
\includegraphics[width=\columnwidth]{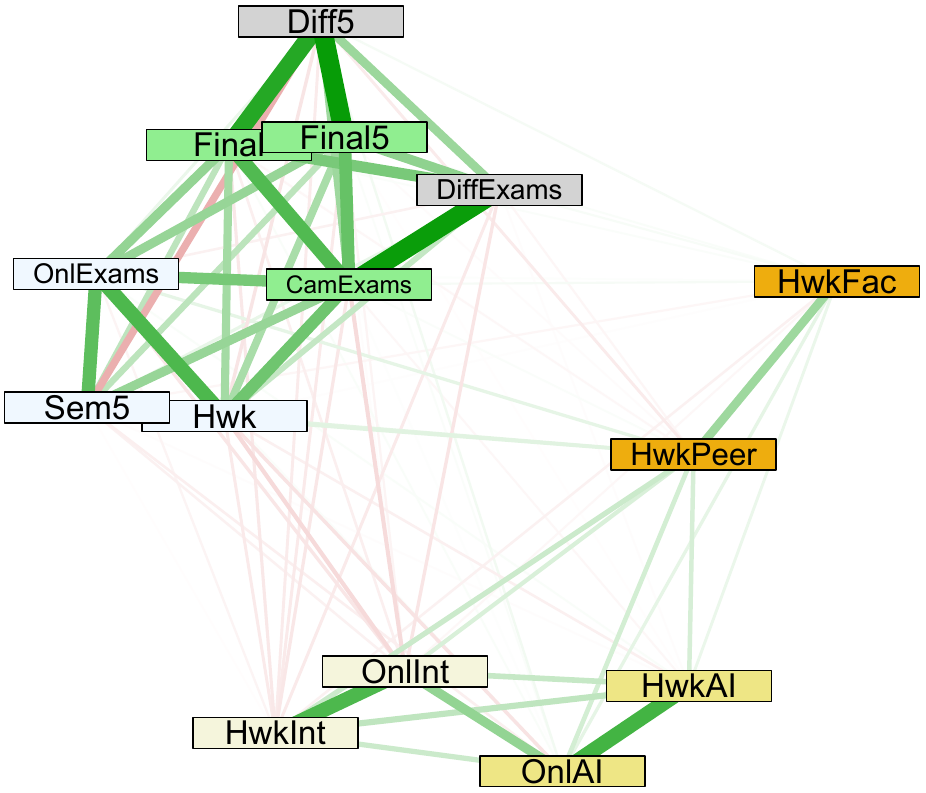}
\end{center}
\caption{Fruchterman-Reingold~\citep{fruchterman1991,qgraph} representation of all correlations between the variables in Table~\ref{tab:variables} for all students.}
\label{fig:correlations}
\end{figure}

Figure~\ref{fig:correlations} shows a Fruchterman-Reingold~\citep{fruchterman1991,qgraph} representation of the correlation matrix between the variables. Indicated in light blue are the online, unsupervised assessments, in green the on-campus, supervised assessments, and in gray the differences in scores between selected subsets of assessments. The percentages of AI-usage are indicated in beige, usage of internet resources in yellow, and discussions with humans in orange. Green edges denote positive correlations, red edges negative correlations, and the thickness their absolute strength. Due to the force-directed nature of Fruchterman-Reingold graphs, closely correlated vertices tend to cluster, while unrelated vertices tend to be further apart from each other.

It is apparent that the scores achieved online and those achieved under supervision each cluster together, but both are disconnected from the survey answers. Taking \attr{Final} as a proxy for learning success in the course, after discarding the derived variable \attr{DiffExams} and all variables related to the five duplicate problems, in a linear regression of all remaining assessment and survey variables, only \attr{CamExam} emerges as a statistically significant predictor of the final exam score ($p\approx7.3\cdot10^{-16}$); \attr{OnlExams} comes close to statistical significance with $p=0.05$. Table~\ref{tab:linreg} shows the outcome of this linear regression, which has a week correlation of $R^2=0.36$.

\begin{table}
\caption{Linear regression for \attr{Final}.\label{tab:linreg}}
\begin{ruledtabular}
\begin{tabular}{lrrrr}
			&Estimate&Std. Dev.&	$t$&	$p$\\\hline
\attr{Hwk}			&$-0.08$	&$0.13$		&$-0.6$		&$0.55$\\
\attr{OnlExams}		&$0.33$	&$0.17$		&$1.96$		&$0.05$\\
\attr{CamExams}	&$0.69$	&$0.08$		&$8.74$		&$7.31\cdot10^{-16}$\\
\attr{HwkAI}		&$-0.06$	&$0.09$		&$-0.68$		&$0.50$\\
\attr{HwkInt}		&$-0.05$	&$0.07$		&$-0.70$		&$0.49$\\
\attr{HwkPeer}		&$-0.04$	&$0.05$		&$-0.78$		&$0.44$\\
\attr{HwkFac}		&$0.01$	&$0.07$		&$0.20$		&$0.84$\\
\attr{OnlAI}		&$0.06$	&$0.12$		&$0.46$		&$0.65$\\
\attr{OnlInt}		&$0.05$	&$0.07$		&$0.74$		&$0.46$\\
\end{tabular}
\end{ruledtabular}
\end{table}

\begin{figure*}
\begin{center}
\includegraphics[width=0.96\columnwidth]{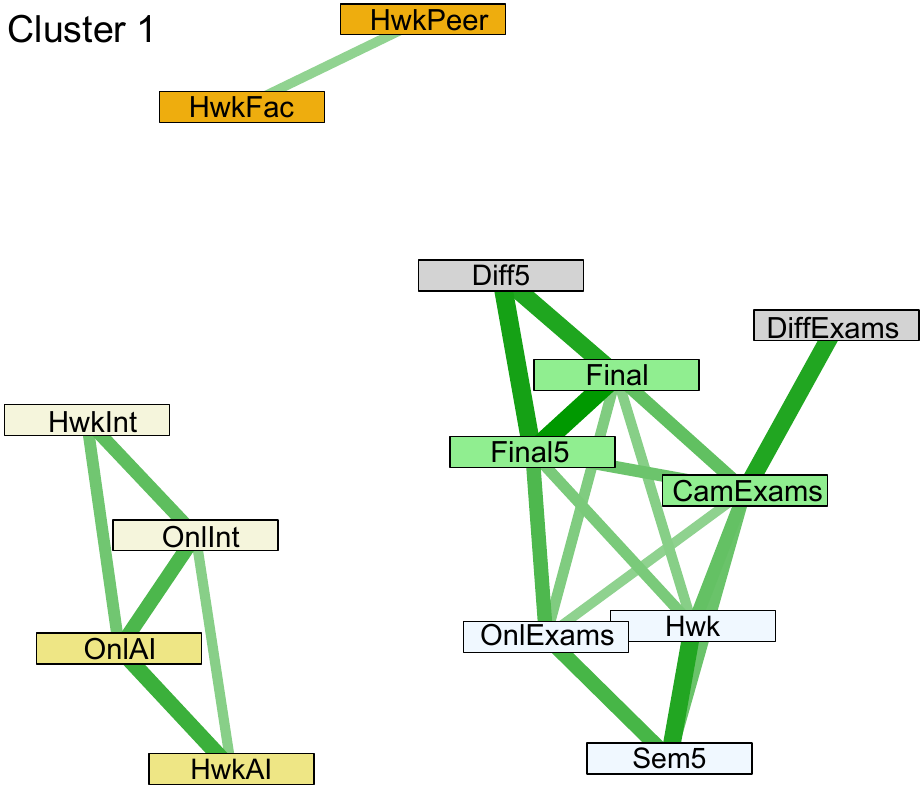}\qquad
\includegraphics[width=0.96\columnwidth]{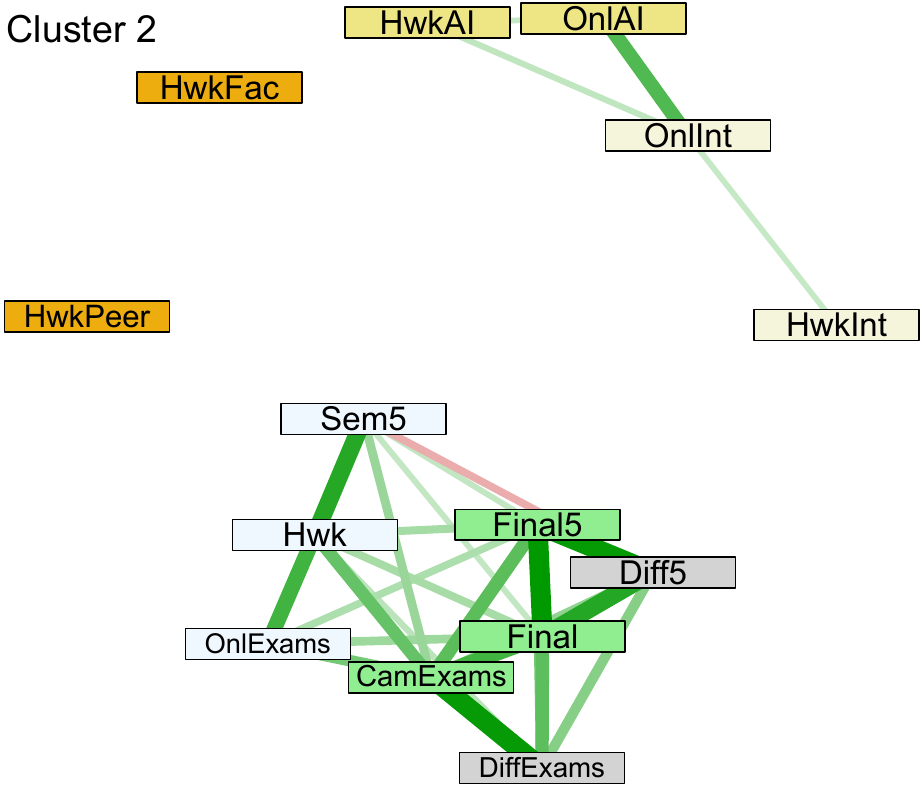}

\vspace*{5mm}

\includegraphics[width=0.96\columnwidth]{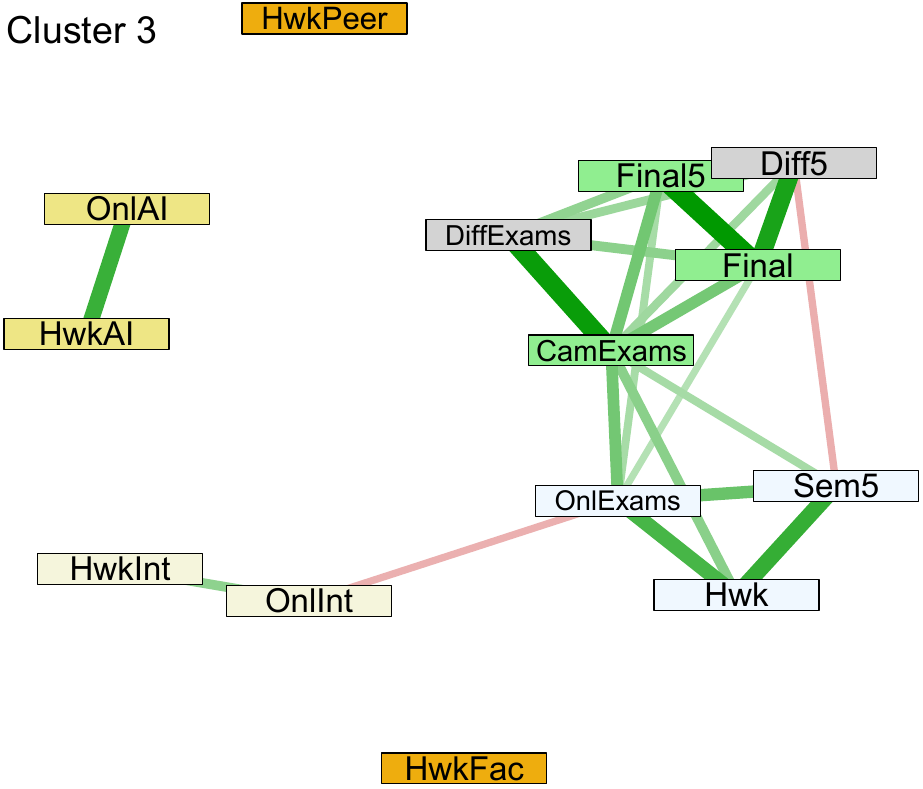}\qquad
\includegraphics[width=0.96\columnwidth]{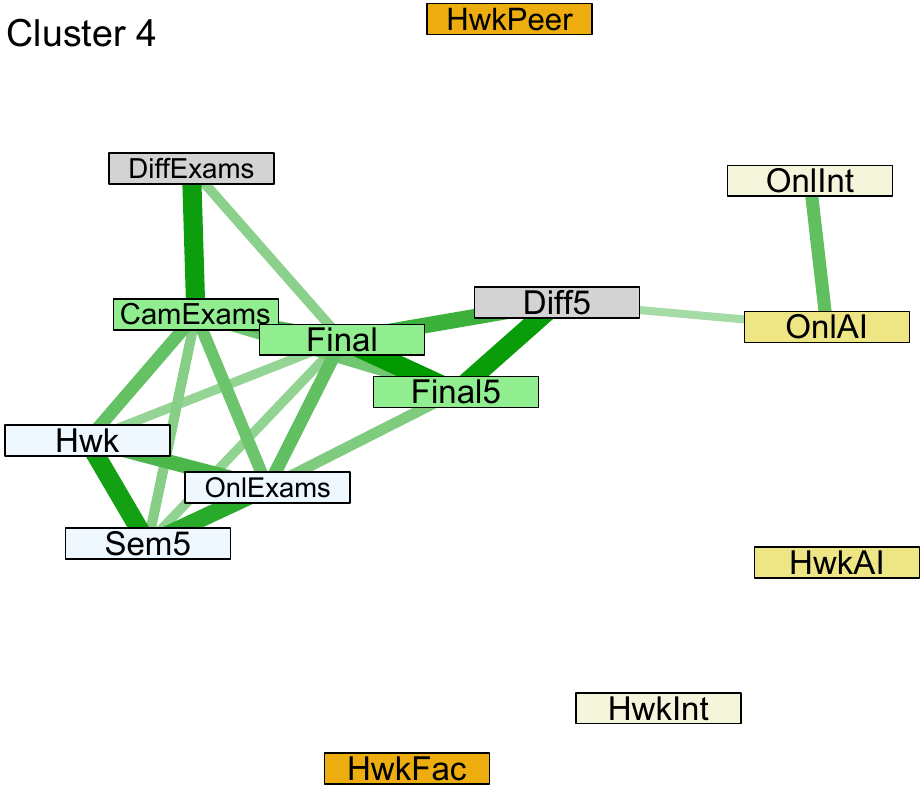}
\end{center}
\caption{Fruchterman-Reingold~\citep{fruchterman1991,qgraph} representation of the statistically significant correlations ($p<0.05$) between the variables in Table~\ref{tab:variables} for the clusters in Table~\ref{tab:clusters}. Note the rotation and handedness of these representations are random.}
\label{fig:cluster_correlations}
\end{figure*}

Also within the usage classes (Table~\ref{tab:clusters}), there are very few significant correlations between resource usage and assessment performance. Figure~\ref{fig:cluster_correlations} shows the statistically significant correlations between the variables ($p<0.05$).  For the heavy internet users (Cluster~3), the usage of internet resources other than AI during online exams (\attr{OnlInt}) is significantly negatively correlated with the scores on the exams (\attr{OnlExams}) ($r=-0.28$; $p=0.04$), which may indicate that relying on the internet, the students were not able to quickly enough find what they needed to correctly solve the problems, including replacement of the numbers by their values. For the users who made use of all resources everywhere (Cluster~4), the use of AI during online exams (\attr{OnlAI}) is significantly positively related to \attr{Diff5} ($r=0.31$; $p=0.04$); this means that AI-usage during online exams is positively correlated with doing better on the final exam instance of duplicate problems than on their first occurrence in the course.

Notably, within all usage classes, significant correlations between the supervised final exam and the unsupervised assessment components of the course remain. This means that in spite of even the heaviest use of external resources, if not as a significant predictor, unsupervised assessment still retains formative relevance. 

Using \attr{Final} as a proxy for learning success, we also investigated if higher or lower performing quartiles of the students may have benefitted or been harmed by the use of online resources, but we found no difference in that regard between these populations.

\subsection{Item analysis}
The use of external resources is detrimental to the validity of assessments. Table~\ref{tab:itemana} shows the average item parameters for assessment items on homework and unsupervised online exams. The mean reflects the average percentage of correctly solved items, and the point biserial (``pBis'') the discrimination of these items (ranging from $-1$ to $1$ where negative values usually denote invalid assessment items), that is, how well the item distinguishes between students who generally have a good grasp of the concepts and those who do not.
\begin{table}
\caption{Average problem solving mean and average point biserial correlations for homework and online exams by clusters.\label{tab:itemana}}
\begin{ruledtabular}
\begin{tabular}{lrrrr}
&\multicolumn{2}{c}{Homework}&\multicolumn{2}{c}{Online Exams}\\
Cluster&Mean&pBis&Mean&pBis\\
\hline
1&0.87	&0.46&0.82	&0.33\\
2&0.86	&0.45&0.82	&0.30\\
3&0.84	&0.40&0.81	&0.28\\
4&0.85	&0.49&0.83	&0.26\\
\end{tabular}
\end{ruledtabular}
\end{table}

The values only insignificantly differ between different classes of online resource usage patterns. Overall, they are consistent with the low predictive power of these unsupervised assessments, but they still indicate that items can provide feedback to students and instructors.
 
\subsection{Student comments about AI}
Based on the replies to the open-ended question on the survey, many students recognize AI as a valuable tool for assisting in learning, particularly for understanding complex topics and guiding problem-solving. They particularly value its ability to quickly provide information without having to flip through textbook materials or scroll through video. They appreciate AI's ability to provide alternate explanations and solutions, which can be especially helpful when traditional teaching methods fall short. However, there's a consensus that AI should not replace genuine learning and effort. Students suggest that AI's role should be that of an assistant rather than a solution provider, and its usage should be context-dependent. For instance, in major-related courses, students advocate for minimal AI use to ensure a solid understanding of essential concepts. Conversely, in subjects that are less critical for them, they see AI as a more acceptable aid. 

Concerns about academic integrity and the potential for AI to promote laziness and dependency are prominent. Students worry that reliance on AI for problem-solving or essay writing could lead to a superficial understanding of course material and hinder the development of critical thinking skills. They propose a balanced approach, where AI is used judiciously to enhance learning without becoming a crutch. This balance involves using AI for initial guidance or concept clarification while avoiding its use for directly solving assignments or exams. Some students commented on having some of the exams during the semester being in-person as beneficial. Ironically, but of course also very temptingly, based on style considerations, it appears that several students filled out the free-response question using ChatGPT.

Finally, the practicality of regulating AI use in online education is a significant concern. Some students acknowledge that while AI tools like ChatGPT may not be sophisticated enough currently to solve complex academic problems accurately, they could still be misused. Several students find ChatGPT is not yet trustworthy, but expect this to improve in the future. There's an acknowledgment that AI is a part of the evolving educational landscape, and rather than outright banning it, educators should find ways to integrate it responsibly into the curriculum. This integration could involve designing assessments that still require a deep understanding of the material, even with AI assistance, and teaching students how to use AI ethically and effectively as part of their learning toolkit.

\section{Discussion}
Students are making extensive use of external resources when working on unsupervised assessments. On year after becoming available, students have not made the jump from ``traditional'' problem-solving sites to AI. The reasons might be manifold: the free version of GPT (at the time of writing version 3.5) is much less powerful than version~4, which is only available to subscribers. A GPT-subscription costs \$20/month, while for example Chegg costs \$14.95/month with promotion sales for half that price. Students may also be used to the ``traditional'' resources from high school and carry over their habits to college. For questions that contain figures or graphs, on forums, it is sufficient to submit only the text in order to locate the question, while with AI tools, these illustrations need to be described in words~\cite{kortemeyer23ai} (this will change as the multimodal capabilities of these systems develop further). Finally, with ``traditional'' sites, students would find the exact problem with the expected answer, while all Large Language Models still hallucinate. 

A surprising result of this study is the lack of correlations between self-reported resource usage and assessment outcomes. While the discrepancy between the score distributions of online and on-campus assessments (Fig.~\ref{fig:scoredist}) could be explained by the usage of AI and other online resources (Fig.~\ref{fig:usage}), one would have expected a correlation~\cite{kashy2003influence,palazzo2010}: the more resources are used, the higher the discrepancy; this, however, is not the case. This null result may be due to students being worried about punitive measures in spite of the strict research protocol or students underestimating their reliance on external resources; underreporting of academic dishonesty by approximately $1/3^{\mbox{\scriptsize rd}}$ has been reported before~\cite{palazzo2010}. It is also surprising that in spite of all of the external resource usage, a significant correlation remains between unsupervised and supervised assessments; while the best and only statistically significant predictor of the score on the final exam are the scores on the supervised in-semester exams, the other assessments have not lost their formative relevance.

From the results, it is clear that high-stake exams like the final exam cannot be conducted in unsupervised settings. At the moment, the time it takes to look up solutions on the internet may still be a  hinderance to overly relying on those resources (as was also found by the negative correlation between the extent of using the internet and performance on online exams among students who heavily rely on external resources), on the long run, AI will likely be reliable enough that answers to any introductory physics problem, including newly created ones, can be obtained instantaneously.

Simple usage of lockdown browsers such as Respondus~\cite{respondus} or Safe Exam Browser~\cite{seb} are no remedy in an online setting, as students can simply use another machine or their phone to access sites such as ChatGPT~\cite{chatgpt}, Gemini~\cite{gemini} or Chegg~\cite{chegg}. Instead, these lockdown browsers, which limit access to local disks and particular internet sites, are useful in supervised on-campus settings where students use their own devices (on-campus BYOD exams). If for logistical reasons, high-stake exams have to be conducted online, there is no alternative to intrusive proctoring systems that use cameras and microphones.

Students are well-aware of the possible pitfalls associated with AI usage. While they argue that it will be part of their professional lives, they support a balanced approach to its use, in particular over-dependence and over-reliance. They are also are aware of reliability and trustworthiness issues, which agrees with earlier findings regarding students' ability to judge the quality of answers~\cite{shoufan2023exploring,dahlkemper23}. Overall, though, some statements about using these tools for learning purposes may have to be taken with the same grain of salt as the statement about the expert solution in Fig.~\ref{fig:forum} being ``designed to help students [\ldots] learn core concepts;'' students may believe these statements, but still not act accordingly~\cite{gray09}. Remarks about courses that are ``not critical for them'' suggest that non-sanctioned external resource usage may be particularly strong when the goal is simply to pass the course~\cite{lin}.

\section{Limitations}
Students self-selected into this study, and students who are self-aware that they are using external resources in non-constructive ways may have refrained from participating. As the survey and consent form were administered at the end of the semester, students who dropped out of the course earlier were not considered.

\section{Conclusion}
Our findings reveal that despite the emergence of LLMs, students predominantly rely on traditional internet resources for unsupervised quizzes. However, this reliance does not significantly affect supervised assessments.

Our data indicates that the unsupervised online exams and supervised on-campus exams yield markedly different outcomes, presumably due to the use of external resources in unsupervised settings. Interestingly, there's no strong correlation between self-reported resource usage and exam performance, suggesting other factors at play or potential underreporting of resource usage. Despite heavy resource use, significant correlations between supervised and unsupervised assessments persist, indicating that unsupervised assessments, while having no significant predictive properties, retain formative value.

While students advocate for a balanced approach to AI use, emphasizing its role as an assistant rather than a solution provider, the study underscores the necessity of carrying out high-stakes exams in supervised settings to ensure academic integrity.

Only one quarter of our survey respondents claims to have utilized at least some minimal form of AI tools in the solution of their homework problems.  AI systems, though, are becoming exponentially more competent.  They will rapidly penetrate all education settings.  Our present results do not show significant causal effects on student learning success from using AI tools.  In this sense our study was conducted perhaps a bit early.  But it is clear, nevertheless, that all of us need to rethink our course offerings, and particularly our assessment tools, due to the rise of AI tools.  Now is the time to help shape AI environments into the perfect one-on-one tutor for students, similar to what Feynman envisioned, instead of a means to avoid learning physics.

\begin{acknowledgments}
We would like to thank the students who participated in this study. We would also like to thank Christine Kor\-te\-meyer for her help with the manuscript.
\end{acknowledgments}

\bibliography{OnlineCourses}

\begin{thebibliography}{33}%
\makeatletter
\providecommand \@ifxundefined [1]{%
 \@ifx{#1\undefined}
}%
\providecommand \@ifnum [1]{%
 \ifnum #1\expandafter \@firstoftwo
 \else \expandafter \@secondoftwo
 \fi
}%
\providecommand \@ifx [1]{%
 \ifx #1\expandafter \@firstoftwo
 \else \expandafter \@secondoftwo
 \fi
}%
\providecommand \natexlab [1]{#1}%
\providecommand \enquote  [1]{``#1''}%
\providecommand \bibnamefont  [1]{#1}%
\providecommand \bibfnamefont [1]{#1}%
\providecommand \citenamefont [1]{#1}%
\providecommand \href@noop [0]{\@secondoftwo}%
\providecommand \href [0]{\begingroup \@sanitize@url \@href}%
\providecommand \@href[1]{\@@startlink{#1}\@@href}%
\providecommand \@@href[1]{\endgroup#1\@@endlink}%
\providecommand \@sanitize@url [0]{\catcode `\\12\catcode `\$12\catcode
  `\&12\catcode `\#12\catcode `\^12\catcode `\_12\catcode `\%12\relax}%
\providecommand \@@startlink[1]{}%
\providecommand \@@endlink[0]{}%
\providecommand \url  [0]{\begingroup\@sanitize@url \@url }%
\providecommand \@url [1]{\endgroup\@href {#1}{\urlprefix }}%
\providecommand \urlprefix  [0]{URL }%
\providecommand \Eprint [0]{\href }%
\providecommand \doibase [0]{https://doi.org/}%
\providecommand \selectlanguage [0]{\@gobble}%
\providecommand \bibinfo  [0]{\@secondoftwo}%
\providecommand \bibfield  [0]{\@secondoftwo}%
\providecommand \translation [1]{[#1]}%
\providecommand \BibitemOpen [0]{}%
\providecommand \bibitemStop [0]{}%
\providecommand \bibitemNoStop [0]{.\EOS\space}%
\providecommand \EOS [0]{\spacefactor3000\relax}%
\providecommand \BibitemShut  [1]{\csname bibitem#1\endcsname}%
\let\auto@bib@innerbib\@empty
\bibitem [{\citenamefont {Feynman}\ \emph {et~al.}(1965)\citenamefont
  {Feynman}, \citenamefont {Leighton},\ and\ \citenamefont
  {Sands}}]{feynman1965m}%
  \BibitemOpen
  \bibfield  {author} {\bibinfo {author} {\bibfnamefont {R.}~\bibnamefont
  {Feynman}}, \bibinfo {author} {\bibfnamefont {R.}~\bibnamefont {Leighton}},\
  and\ \bibinfo {author} {\bibfnamefont {M.}~\bibnamefont {Sands}},\
  }\href@noop {} {\emph {\bibinfo {title} {The Feynman Lectures on Physics}}}\
  (\bibinfo  {publisher} {Addison Wesley},\ \bibinfo {address} {Reading, MA},\
  \bibinfo {year} {1965})\BibitemShut {NoStop}%
\bibitem [{\citenamefont {Kortemeyer}\ and\ \citenamefont
  {Bauer}(1999)}]{kortemeyer1999}%
  \BibitemOpen
  \bibfield  {author} {\bibinfo {author} {\bibfnamefont {G.}~\bibnamefont
  {Kortemeyer}}\ and\ \bibinfo {author} {\bibfnamefont {W.}~\bibnamefont
  {Bauer}},\ }\bibfield  {title} {\bibinfo {title} {Multimedia collaborative
  content creation (mc3): The msu lecture online system},\ }\href@noop {}
  {\bibfield  {journal} {\bibinfo  {journal} {Journal of Engineering
  Education}\ }\textbf {\bibinfo {volume} {88}},\ \bibinfo {pages} {421}
  (\bibinfo {year} {1999})}\BibitemShut {NoStop}%
\bibitem [{\citenamefont {Bauer}\ and\ \citenamefont
  {Westfall}(2023)}]{bauer2023}%
  \BibitemOpen
  \bibfield  {author} {\bibinfo {author} {\bibfnamefont {W.}~\bibnamefont
  {Bauer}}\ and\ \bibinfo {author} {\bibfnamefont {G.~D.}\ \bibnamefont
  {Westfall}},\ }\href
  {https://www.mheducation.com/highered/product/university-physics-modern-physics-bauer-westfall/M9781266672620.html}
  {\emph {\bibinfo {title} {University physics with modern physics (ebook)}}}\
  (\bibinfo  {publisher} {(McGraw-Hill Higher Education)},\ \bibinfo {address}
  {New York, NY},\ \bibinfo {year} {2023})\BibitemShut {NoStop}%
\bibitem [{\citenamefont {Kortemeyer}(2014)}]{kortemeyer2014onl}%
  \BibitemOpen
  \bibfield  {author} {\bibinfo {author} {\bibfnamefont {G.}~\bibnamefont
  {Kortemeyer}},\ }\bibfield  {title} {\bibinfo {title} {Over two decades of
  blended and online physics courses at {Michigan State University}},\
  }\href@noop {} {\bibfield  {journal} {\bibinfo  {journal} {eleed}\ }\textbf
  {\bibinfo {volume} {10}} (\bibinfo {year} {2014})}\BibitemShut {NoStop}%
\bibitem [{\citenamefont {Russell}(1997)}]{russell1997no}%
  \BibitemOpen
  \bibfield  {author} {\bibinfo {author} {\bibfnamefont {T.~L.}\ \bibnamefont
  {Russell}},\ }\href@noop {} {\emph {\bibinfo {title} {The ``No Significant
  Difference'' phenomenon as reported in 248 research reports, summaries, and
  papers}}}\ (\bibinfo  {publisher} {North Carolina State University},\
  \bibinfo {year} {1997})\BibitemShut {NoStop}%
\bibitem [{\citenamefont {Cavanaugh}\ and\ \citenamefont
  {Jacquemin}(2015)}]{cavanaugh2015large}%
  \BibitemOpen
  \bibfield  {author} {\bibinfo {author} {\bibfnamefont {J.}~\bibnamefont
  {Cavanaugh}}\ and\ \bibinfo {author} {\bibfnamefont {S.~J.}\ \bibnamefont
  {Jacquemin}},\ }\bibfield  {title} {\bibinfo {title} {A large sample
  comparison of grade based student learning outcomes in online vs.
  face-to-face courses},\ }\href@noop {} {\bibfield  {journal} {\bibinfo
  {journal} {Online learning}\ }\textbf {\bibinfo {volume} {19}} (\bibinfo
  {year} {2015})}\BibitemShut {NoStop}%
\bibitem [{\citenamefont {Bergeler}\ and\ \citenamefont
  {Read}(2021)}]{bergeler2021comparing}%
  \BibitemOpen
  \bibfield  {author} {\bibinfo {author} {\bibfnamefont {E.}~\bibnamefont
  {Bergeler}}\ and\ \bibinfo {author} {\bibfnamefont {M.~F.}\ \bibnamefont
  {Read}},\ }\bibfield  {title} {\bibinfo {title} {Comparing learning outcomes
  and satisfaction of an online algebra-based physics course with a
  face-to-face course},\ }\href@noop {} {\bibfield  {journal} {\bibinfo
  {journal} {Journal of Science Education and Technology}\ }\textbf {\bibinfo
  {volume} {30}},\ \bibinfo {pages} {97} (\bibinfo {year} {2021})}\BibitemShut
  {NoStop}%
\bibitem [{\citenamefont {Kortemeyer}\ \emph {et~al.}(2022)\citenamefont
  {Kortemeyer}, \citenamefont {Bauer},\ and\ \citenamefont
  {Fisher}}]{kortemeyer22hybrid}%
  \BibitemOpen
  \bibfield  {author} {\bibinfo {author} {\bibfnamefont {G.}~\bibnamefont
  {Kortemeyer}}, \bibinfo {author} {\bibfnamefont {W.}~\bibnamefont {Bauer}},\
  and\ \bibinfo {author} {\bibfnamefont {W.}~\bibnamefont {Fisher}},\
  }\bibfield  {title} {\bibinfo {title} {Hybrid teaching: A tale of two
  populations},\ }\href@noop {} {\bibfield  {journal} {\bibinfo  {journal}
  {Physical Review Physics Education Research}\ }\textbf {\bibinfo {volume}
  {18}},\ \bibinfo {pages} {020130} (\bibinfo {year} {2022})}\BibitemShut
  {NoStop}%
\bibitem [{\citenamefont {Kortemeyer}\ \emph {et~al.}(2023)\citenamefont
  {Kortemeyer}, \citenamefont {Kortemeyer},\ and\ \citenamefont
  {Bauer}}]{kortemeyer2023taking}%
  \BibitemOpen
  \bibfield  {author} {\bibinfo {author} {\bibfnamefont {G.}~\bibnamefont
  {Kortemeyer}}, \bibinfo {author} {\bibfnamefont {C.}~\bibnamefont
  {Kortemeyer}},\ and\ \bibinfo {author} {\bibfnamefont {W.}~\bibnamefont
  {Bauer}},\ }\bibfield  {title} {\bibinfo {title} {Taking introductory physics
  in studio, lecture, or online format: What difference does it make in
  subsequent courses, and for whom?},\ }\href
  {https://doi.org/10.1103/PhysRevPhysEducRes.19.020148} {\bibfield  {journal}
  {\bibinfo  {journal} {Phys. Rev. Phys. Educ. Res.}\ }\textbf {\bibinfo
  {volume} {19}},\ \bibinfo {pages} {020148} (\bibinfo {year}
  {2023})}\BibitemShut {NoStop}%
\bibitem [{\citenamefont {Ruggieri}(2020)}]{ruggieri2020students}%
  \BibitemOpen
  \bibfield  {author} {\bibinfo {author} {\bibfnamefont {C.}~\bibnamefont
  {Ruggieri}},\ }\bibfield  {title} {\bibinfo {title} {Students' use and
  perception of textbooks and online resources in introductory physics},\
  }\href@noop {} {\bibfield  {journal} {\bibinfo  {journal} {Physical Review
  Physics Education Research}\ }\textbf {\bibinfo {volume} {16}},\ \bibinfo
  {pages} {020123} (\bibinfo {year} {2020})}\BibitemShut {NoStop}%
\bibitem [{\citenamefont {Chegg}(2023)}]{chegg}%
  \BibitemOpen
  \bibfield  {author} {\bibinfo {author} {\bibnamefont {Chegg}},\ }\href@noop
  {} {\bibinfo {title} {Chegg}},\ \bibinfo {howpublished}
  {\url{https://www.chegg.com}} (\bibinfo {year} {accessed December
  2023})\BibitemShut {NoStop}%
\bibitem [{\citenamefont {Lancaster}\ and\ \citenamefont
  {Cotarlan}(2021)}]{lancaster2021contract}%
  \BibitemOpen
  \bibfield  {author} {\bibinfo {author} {\bibfnamefont {T.}~\bibnamefont
  {Lancaster}}\ and\ \bibinfo {author} {\bibfnamefont {C.}~\bibnamefont
  {Cotarlan}},\ }\bibfield  {title} {\bibinfo {title} {Contract cheating by
  stem students through a file sharing website: a covid-19 pandemic
  perspective},\ }\href@noop {} {\bibfield  {journal} {\bibinfo  {journal}
  {International Journal for Educational Integrity}\ }\textbf {\bibinfo
  {volume} {17}},\ \bibinfo {pages} {1} (\bibinfo {year} {2021})}\BibitemShut
  {NoStop}%
\bibitem [{\citenamefont {Kortemeyer}\ \emph {et~al.}(2008)\citenamefont
  {Kortemeyer}, \citenamefont {Kashy}, \citenamefont {Benenson},\ and\
  \citenamefont {Bauer}}]{kortemeyer08}%
  \BibitemOpen
  \bibfield  {author} {\bibinfo {author} {\bibfnamefont {G.}~\bibnamefont
  {Kortemeyer}}, \bibinfo {author} {\bibfnamefont {E.}~\bibnamefont {Kashy}},
  \bibinfo {author} {\bibfnamefont {W.}~\bibnamefont {Benenson}},\ and\
  \bibinfo {author} {\bibfnamefont {W.}~\bibnamefont {Bauer}},\ }\bibfield
  {title} {\bibinfo {title} {Experiences using the open-source learning content
  management and assessment system {LON-CAPA} in introductory physics
  courses},\ }\href@noop {} {\bibfield  {journal} {\bibinfo  {journal} {Am. J.
  Phys}\ }\textbf {\bibinfo {volume} {76}},\ \bibinfo {pages} {438} (\bibinfo
  {year} {2008})}\BibitemShut {NoStop}%
\bibitem [{\citenamefont {Busch}(2017)}]{busch2017one}%
  \BibitemOpen
  \bibfield  {author} {\bibinfo {author} {\bibfnamefont {H.}~\bibnamefont
  {Busch}},\ }\bibfield  {title} {\bibinfo {title} {One method for inhibiting
  the copying of online homework},\ }\href@noop {} {\bibfield  {journal}
  {\bibinfo  {journal} {The Physics Teacher}\ }\textbf {\bibinfo {volume}
  {55}},\ \bibinfo {pages} {422} (\bibinfo {year} {2017})}\BibitemShut
  {NoStop}%
\bibitem [{\citenamefont {{OpenAI}}(2023)}]{chatgpt}%
  \BibitemOpen
  \bibfield  {author} {\bibinfo {author} {\bibnamefont {{OpenAI}}},\
  }\href@noop {} {\bibinfo {title} {{ChatGPT}}},\ \bibinfo {howpublished}
  {\url{https://chat.openai.com/}} (\bibinfo {year} {accessed December
  2023})\BibitemShut {NoStop}%
\bibitem [{\citenamefont {{Google}}(2023{\natexlab{a}})}]{bard}%
  \BibitemOpen
  \bibfield  {author} {\bibinfo {author} {\bibnamefont {{Google}}},\
  }\href@noop {} {\bibinfo {title} {{Google Bard}}},\ \bibinfo {howpublished}
  {\url{https://bard.google.com/}} (\bibinfo {year} {accessed December
  2023}{\natexlab{a}})\BibitemShut {NoStop}%
\bibitem [{\citenamefont {Prakash}(2023)}]{forbes}%
  \BibitemOpen
  \bibfield  {author} {\bibinfo {author} {\bibfnamefont {P.}~\bibnamefont
  {Prakash}},\ }\bibfield  {title} {\bibinfo {title} {Chegg's shares tumbled
  nearly 50\% after the edtech company said its customers are using {ChatGPT}
  instead of paying for its study tools},\ }\href@noop {} {\bibfield  {journal}
  {\bibinfo  {journal} {Forbes}\ }\textbf {\bibinfo {volume} {May}} (\bibinfo
  {year} {2023})}\BibitemShut {NoStop}%
\bibitem [{\citenamefont {Paresh}(2023)}]{wired}%
  \BibitemOpen
  \bibfield  {author} {\bibinfo {author} {\bibfnamefont {D.}~\bibnamefont
  {Paresh}},\ }\bibfield  {title} {\bibinfo {title} {Chegg embraced ai.
  {ChatGPT} ate its lunch anyway},\ }\href@noop {} {\bibfield  {journal}
  {\bibinfo  {journal} {Wired}\ }\textbf {\bibinfo {volume} {June}} (\bibinfo
  {year} {2023})}\BibitemShut {NoStop}%
\bibitem [{\citenamefont {Kortemeyer}(2023)}]{kortemeyer23ai}%
  \BibitemOpen
  \bibfield  {author} {\bibinfo {author} {\bibfnamefont {G.}~\bibnamefont
  {Kortemeyer}},\ }\bibfield  {title} {\bibinfo {title} {Could an
  artificial-intelligence agent pass an introductory physics course?},\ }\href
  {https://doi.org/10.1103/PhysRevPhysEducRes.19.010132} {\bibfield  {journal}
  {\bibinfo  {journal} {Phys. Rev. Phys. Educ. Res.}\ }\textbf {\bibinfo
  {volume} {19}},\ \bibinfo {pages} {010132} (\bibinfo {year}
  {2023})}\BibitemShut {NoStop}%
\bibitem [{\citenamefont {Laverty}\ \emph {et~al.}(2012)\citenamefont
  {Laverty}, \citenamefont {Bauer}, \citenamefont {Kortemeyer},\ and\
  \citenamefont {Westfall}}]{laverty12b}%
  \BibitemOpen
  \bibfield  {author} {\bibinfo {author} {\bibfnamefont {J.~T.}\ \bibnamefont
  {Laverty}}, \bibinfo {author} {\bibfnamefont {W.}~\bibnamefont {Bauer}},
  \bibinfo {author} {\bibfnamefont {G.}~\bibnamefont {Kortemeyer}},\ and\
  \bibinfo {author} {\bibfnamefont {G.}~\bibnamefont {Westfall}},\ }\bibfield
  {title} {\bibinfo {title} {Want to reduce guessing and cheating while making
  students happier? give more exams!},\ }\href@noop {} {\bibfield  {journal}
  {\bibinfo  {journal} {Phys. Teach.}\ }\textbf {\bibinfo {volume} {50}},\
  \bibinfo {pages} {540} (\bibinfo {year} {2012})}\BibitemShut {NoStop}%
\bibitem [{\citenamefont {{R Core Team}}(2021)}]{rproject}%
  \BibitemOpen
  \bibfield  {author} {\bibinfo {author} {\bibnamefont {{R Core Team}}},\
  }\href {https://www.R-project.org/} {\emph {\bibinfo {title} {R: A Language
  and Environment for Statistical Computing}}},\ \bibinfo {organization} {R
  Foundation for Statistical Computing},\ \bibinfo {address} {Vienna, Austria}
  (\bibinfo {year} {2021})\BibitemShut {NoStop}%
\bibitem [{\citenamefont {Epskamp}\ \emph {et~al.}(2012)\citenamefont
  {Epskamp}, \citenamefont {Cramer}, \citenamefont {Waldorp}, \citenamefont
  {Schmittmann},\ and\ \citenamefont {Borsboom}}]{qgraph}%
  \BibitemOpen
  \bibfield  {author} {\bibinfo {author} {\bibfnamefont {S.}~\bibnamefont
  {Epskamp}}, \bibinfo {author} {\bibfnamefont {A.~O.~J.}\ \bibnamefont
  {Cramer}}, \bibinfo {author} {\bibfnamefont {L.~J.}\ \bibnamefont {Waldorp}},
  \bibinfo {author} {\bibfnamefont {V.~D.}\ \bibnamefont {Schmittmann}},\ and\
  \bibinfo {author} {\bibfnamefont {D.}~\bibnamefont {Borsboom}},\ }\bibfield
  {title} {\bibinfo {title} {{qgraph}: Network visualizations of relationships
  in psychometric data},\ }\href {http://www.jstatsoft.org/v48/i04/} {\bibfield
   {journal} {\bibinfo  {journal} {Journal of Statistical Software}\ }\textbf
  {\bibinfo {volume} {48}},\ \bibinfo {pages} {1} (\bibinfo {year}
  {2012})}\BibitemShut {NoStop}%
\bibitem [{\citenamefont {Willse}(2014)}]{ctt}%
  \BibitemOpen
  \bibfield  {author} {\bibinfo {author} {\bibfnamefont {J.~T.}\ \bibnamefont
  {Willse}},\ }\href@noop {} {\bibinfo {title} {ct.: An r package for classical
  test theory functions}} (\bibinfo {year} {2014})\BibitemShut {NoStop}%
\bibitem [{\citenamefont {Fruchterman}\ and\ \citenamefont
  {Reingold}(1991)}]{fruchterman1991}%
  \BibitemOpen
  \bibfield  {author} {\bibinfo {author} {\bibfnamefont {T.~M.}\ \bibnamefont
  {Fruchterman}}\ and\ \bibinfo {author} {\bibfnamefont {E.~M.}\ \bibnamefont
  {Reingold}},\ }\bibfield  {title} {\bibinfo {title} {Graph drawing by
  force-directed placement},\ }\href@noop {} {\bibfield  {journal} {\bibinfo
  {journal} {Software: Practice and experience}\ }\textbf {\bibinfo {volume}
  {21}},\ \bibinfo {pages} {1129} (\bibinfo {year} {1991})}\BibitemShut
  {NoStop}%
\bibitem [{\citenamefont {Kashy}\ \emph {et~al.}(2003)\citenamefont {Kashy},
  \citenamefont {Albertelli}, \citenamefont {Bauer}, \citenamefont {Kashy},\
  and\ \citenamefont {Thoennessen}}]{kashy2003influence}%
  \BibitemOpen
  \bibfield  {author} {\bibinfo {author} {\bibfnamefont {D.~A.}\ \bibnamefont
  {Kashy}}, \bibinfo {author} {\bibfnamefont {G.}~\bibnamefont {Albertelli}},
  \bibinfo {author} {\bibfnamefont {W.}~\bibnamefont {Bauer}}, \bibinfo
  {author} {\bibfnamefont {E.}~\bibnamefont {Kashy}},\ and\ \bibinfo {author}
  {\bibfnamefont {M.}~\bibnamefont {Thoennessen}},\ }\bibfield  {title}
  {\bibinfo {title} {Influence of nonmoderated and moderated discussion sites
  on student success},\ }\href@noop {} {\bibfield  {journal} {\bibinfo
  {journal} {Journal of Asynchronous Learning Networks}\ }\textbf {\bibinfo
  {volume} {7}},\ \bibinfo {pages} {31} (\bibinfo {year} {2003})}\BibitemShut
  {NoStop}%
\bibitem [{\citenamefont {Palazzo}\ \emph {et~al.}(2010)\citenamefont
  {Palazzo}, \citenamefont {Lee}, \citenamefont {Warnakulasooriya},\ and\
  \citenamefont {Pritchard}}]{palazzo2010}%
  \BibitemOpen
  \bibfield  {author} {\bibinfo {author} {\bibfnamefont {D.~J.}\ \bibnamefont
  {Palazzo}}, \bibinfo {author} {\bibfnamefont {Y.-J.}\ \bibnamefont {Lee}},
  \bibinfo {author} {\bibfnamefont {R.}~\bibnamefont {Warnakulasooriya}},\ and\
  \bibinfo {author} {\bibfnamefont {D.~E.}\ \bibnamefont {Pritchard}},\
  }\bibfield  {title} {\bibinfo {title} {Patterns, correlates, and reduction of
  homework copying},\ }\href@noop {} {\bibfield  {journal} {\bibinfo  {journal}
  {Physical Review Special Topics -- Physics Education Research}\ }\textbf
  {\bibinfo {volume} {6}},\ \bibinfo {pages} {010104} (\bibinfo {year}
  {2010})}\BibitemShut {NoStop}%
\bibitem [{\citenamefont {{Respondus}}(2023)}]{respondus}%
  \BibitemOpen
  \bibfield  {author} {\bibinfo {author} {\bibnamefont {{Respondus}}},\
  }\href@noop {} {\bibinfo {title} {{Respondus Browser}}},\ \bibinfo
  {howpublished} {\url{https://web.respondus.com/he/lockdownbrowser/}}
  (\bibinfo {year} {accessed December 2023})\BibitemShut {NoStop}%
\bibitem [{\citenamefont {{SEB Alliance, ETH Zurich}}(2023)}]{seb}%
  \BibitemOpen
  \bibfield  {author} {\bibinfo {author} {\bibnamefont {{SEB Alliance, ETH
  Zurich}}},\ }\href@noop {} {\bibinfo {title} {{Save Exam Browser}}},\
  \bibinfo {howpublished} {\url{https://www.safeexambrowser.org/}} (\bibinfo
  {year} {accessed December 2023})\BibitemShut {NoStop}%
\bibitem [{\citenamefont {{Google}}(2023{\natexlab{b}})}]{gemini}%
  \BibitemOpen
  \bibfield  {author} {\bibinfo {author} {\bibnamefont {{Google}}},\
  }\href@noop {} {\bibinfo {title} {{Gemini}}},\ \bibinfo {howpublished}
  {\url{https://deepmind.google/technologies/gemini/}} (\bibinfo {year}
  {accessed December 2023}{\natexlab{b}})\BibitemShut {NoStop}%
\bibitem [{\citenamefont {Shoufan}(2023)}]{shoufan2023exploring}%
  \BibitemOpen
  \bibfield  {author} {\bibinfo {author} {\bibfnamefont {A.}~\bibnamefont
  {Shoufan}},\ }\bibfield  {title} {\bibinfo {title} {Exploring
  students'perceptions of {CHATGPT}: Thematic analysis and follow-up survey},\
  }\href@noop {} {\bibfield  {journal} {\bibinfo  {journal} {IEEE Access}\ }
  (\bibinfo {year} {2023})}\BibitemShut {NoStop}%
\bibitem [{\citenamefont {Dahlkemper}\ \emph {et~al.}(2023)\citenamefont
  {Dahlkemper}, \citenamefont {Lahme},\ and\ \citenamefont
  {Klein}}]{dahlkemper23}%
  \BibitemOpen
  \bibfield  {author} {\bibinfo {author} {\bibfnamefont {M.~N.}\ \bibnamefont
  {Dahlkemper}}, \bibinfo {author} {\bibfnamefont {S.~Z.}\ \bibnamefont
  {Lahme}},\ and\ \bibinfo {author} {\bibfnamefont {P.}~\bibnamefont {Klein}},\
  }\bibfield  {title} {\bibinfo {title} {How do physics students evaluate
  artificial intelligence responses on comprehension questions? a study on the
  perceived scientific accuracy and linguistic quality of {ChatGPT}},\ }\href
  {https://doi.org/10.1103/PhysRevPhysEducRes.19.010142} {\bibfield  {journal}
  {\bibinfo  {journal} {Phys. Rev. Phys. Educ. Res.}\ }\textbf {\bibinfo
  {volume} {19}},\ \bibinfo {pages} {010142} (\bibinfo {year}
  {2023})}\BibitemShut {NoStop}%
\bibitem [{\citenamefont {Gray}\ \emph {et~al.}(2008)\citenamefont {Gray},
  \citenamefont {Adams}, \citenamefont {Wieman},\ and\ \citenamefont
  {Perkins}}]{gray09}%
  \BibitemOpen
  \bibfield  {author} {\bibinfo {author} {\bibfnamefont {K.~E.}\ \bibnamefont
  {Gray}}, \bibinfo {author} {\bibfnamefont {W.~K.}\ \bibnamefont {Adams}},
  \bibinfo {author} {\bibfnamefont {C.~E.}\ \bibnamefont {Wieman}},\ and\
  \bibinfo {author} {\bibfnamefont {K.~K.}\ \bibnamefont {Perkins}},\
  }\bibfield  {title} {\bibinfo {title} {Students know what physicists believe,
  but they don\char39{}t agree: A study using the class survey},\ }\href
  {https://doi.org/10.1103/PhysRevSTPER.4.020106} {\bibfield  {journal}
  {\bibinfo  {journal} {Phys. Rev. ST Phys. Educ. Res.}\ }\textbf {\bibinfo
  {volume} {4}},\ \bibinfo {pages} {020106} (\bibinfo {year}
  {2008})}\BibitemShut {NoStop}%
\bibitem [{\citenamefont {Lin}(1982)}]{lin}%
  \BibitemOpen
  \bibfield  {author} {\bibinfo {author} {\bibfnamefont {H.}~\bibnamefont
  {Lin}},\ }\bibfield  {title} {\bibinfo {title} {Learning physics vs. passing
  courses},\ }\href@noop {} {\bibfield  {journal} {\bibinfo  {journal} {Phys.
  Teach.}\ }\textbf {\bibinfo {volume} {20}},\ \bibinfo {pages} {151} (\bibinfo
  {year} {1982})}\BibitemShut {NoStop}%
\end{thebibliography}%

\end{document}